# DESI Commissioning Instrument Metrology

Rebecca A. Coles[a], David Brooks[b], Mark Derwent[a], Paul Martini[a], Thomas O'Brien[a], Ashley Ross[a], and Suk Sien Tie[a]

[a]The Ohio State University, Department of Astronomy, Columbus Ohio 43210, USA
[b]Department of Physics & Astronomy, University College London, Gower Street, London, WC1E 6BT, UK

## ABSTRACT

The Dark Energy Spectroscopic Instrument (DESI) is under construction to measure the expansion history of the Universe using the Baryon Acoustic Oscillation technique. The spectra of 35 million galaxies and quasars over 14000 sq deg will be measured during the life of the experiment. A new prime focus corrector for the KPNO Mayall telescope will deliver light to 5000 fiber optic positioners. The fibers in turn feed ten broad-band spectrographs. We will describe the methods and results for the commissioning instrument metrology program. The primary goals of this program are to calculate the transformations and further develop the systems that will place fibers within $5\mu$m RMS of the target positions. We will use the commissioning instrument metrology program to measure the absolute three axis Cartesian coordinates of the five CCDs and 22 illuminated fiducials on the commissioning instrument.

**Keywords:** DESI, Commissioning, Metrology

## 1. INTRODUCTION

The Dark Energy Spectroscopic Instrument (DESI) is a ground based experiment that will measure baryon acoustic oscillations (BAO) and redshift-space distortions (RSD) with an ambitious new wide-area redshift survey. DESI consists of ten broad-band and fiber-fed spectrographs that will take up to 5000 simultaneous spectra that range in wavelength from 360 nm to 980 nm. The instrument will be installed at the 4-m Mayall telescope in Kitt Peak, Arizona, accompanied by an optical corrector that boasts a three-degree diameter field of view.[1] Over the course of a five year survey, the DESI instrument will measure the spectra of 35 million galaxies and quasars over 14000 $\deg^2$.

The new, 8-square degree prime focus corrector produces an 0.81m diameter aspheric focal surface with a mean plate scale of 14.19 arcsec mm$^{-1}$. The DESI focal plane system (FPS) has 5000 robotically controlled fiber positioners that will feed ten three-arm spectrographs with resolution $R = \lambda/\Delta\lambda$ between 2000 and 5500.[1] The positioners will align the fibers so that each captures the light of a single target. Interspersed among the fiber positioners are 120 illuminated fiducials (IFs) that provide a fixed reference for the fiber positioners. The FPS will place the tips of all 5000 fibers within $5\mu$m RMS of their nominal target position on the focal surface.

The FPS will use a Fiber View Camera (FVC) to take an image of the fiber plane to measure the position of each fiber within $3\mu$m RMS after the fibers have been moved. By using the positions of the fixed field illuminated fiducials as a reference, the positions of any misplaced fibers can then be corrected based on the information from the FVC. The fiber positioners were designed for rapid reconfiguration with the goal of reconfiguring the focal plane for a new field within one minute. The high-speed reconfiguration is essential to ensuring that the survey can be completed within five years.

## 2. DESI COMMISSIONING INSTRUMENT

The DESI commissioning instrument (CI), will be used to commission the telescope control system, guiding, and the Aetin Optics System that maintains optical alignment. It will be used to characterize the image quality and optical distortions of the new corrector. The CI has five commercial cameras; one at the center of the focal surface and four at the periphery of the field that are aligned with the cardinal directions. There are 22 illuminated fiducials to calculate the transformations and further develop the systems that will place fibers within $5\mu$m RMS of the target positions. We will use the commissioning instrument metrology program to measure the absolute three axis Cartesian coordinates of the five CCDs and 22 illuminated fiducials on the commissioning instrument. Fig. 1 shows a model of the CI. The CI is described in more detail elsewhere in these proceedings.[2]

When the CI is installed in place of the future DESI focal plane system (FPS), the center CCD camera will be aligned with the telescopes optical axis, and the other four cameras will will sit 90 degrees apart and at the same radii as the Guide Focus and Alignment cameras (GFAs) that will sit on the FPS. The GFAs will be used to guide the telescope and focus and align the DESI corrector, and we will use the cameras on the CI for the same purpose. In a similar manner, the CI's illuminated fiducials will allow us to measure the parameters needed such that the fixed illuminated fiducials (IFs) present on the focal plane system will be in focus when viewed by the fiber view camera.

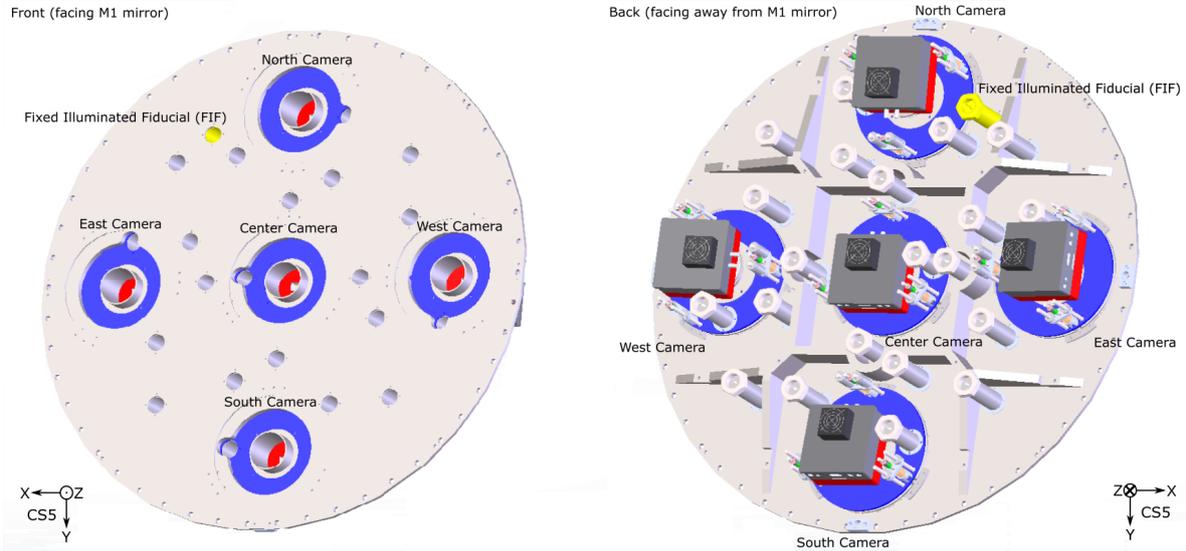

Figure 1. Drawing of the DESI commissioning instrument. Left: the front of the CI, where CS5 Z points towards the primary M1 mirror (out of the page), CS5 X points towards the east camera, and CS5 Y points towards the south camera. The CI contains five SBIG cameras and 22 IFs. The IFs are seated inside of receded tube mounts. Right: the back of the CI, where CS5 Z points into the page. Groups of four IF are placed at various radii that are centered at the center of the CI, which in the CS5 coordinate system will the XYZ position (0,0,0). The center of the CCD of the center camera will be within 2mm of the CS5 Z axis such that we can derive the initial telescope pointing solution and astronomy for the corrector. The North, East, South, and West camera CCDs will be within 2mm of the same radial positions as the GFAs, which is necessary to map the corrector distortions. The IF shown in yellow is the reference IF; it isn't a part of any of the IF radial quadrature, and is used for reference measurements. The blue plates are the the tip/tilt/focus (TTF) mounts that attach the cameras to the CI baseplate, and allow us to make adjustments to camera positions.

The CI will also be used to commission the DESI PlateMaker software to map celestial object locations on the focal plane. The PlateMaker software requires both an accurate model of the distortion pattern created by the primary mirror and optical corrector, and the distortion pattern caused by the optical corrector and FVC optics and imaged onto the FVC CCD. During the PlateMaker commissioning, we will use astrometric data from the five CCDs to improve our model for the DESI corrector. We will then use the DESI CI metrology software to determine the positions of the CI instruments in the coordinate system for the FPS, known as CS5, which aligns the center of the FPS with the center of the M1 mirror. In addition, PlateMaker will predict the pixel position on the FVC of light from the CI IFs. Completion of these key steps with the CI should substantially speed commissioning of the fiber positioning with PlateMaker for the FPS.

## 3. METROLOGY PROCEDURE

There are two types of requirements that the CI must meet: alignment and metrology. The alignment requirements determine how the CI components are placed with respect to each other, as well as their positions in the CS5 coordinate system. For example, the centers of the North, East, South, and West camera CCDs must be within 2mm of the same radial positions as the GFAs. This requirement flows down from the commissioning of the Aetin Optics System. The CCDs and fiducial tips must also be located within $50\mu m$ of the aspheric focal surface, thus each CCD is mounted on an adjustable Tip/Tilt/Focus (TTF) stage to precisely place the CCD on the aspheric focal surface shown in Fig. 2. The metrology requirements are much tighter and stem from the desire to commission many aspects of PlateMaker. There are two metrology requirements, the first being that we must measure the locations of all 22 IFs within $10\mu m$ RMS uncertainty. This is to measure the corrector distortion with the FVC. The second is that we must measure the relative locations of each CCD's pixels and the nearest IF within $5\mu m$ RMS uncertainty. This is to combine the astrometric data with the physical coordinate system of the focal surface.

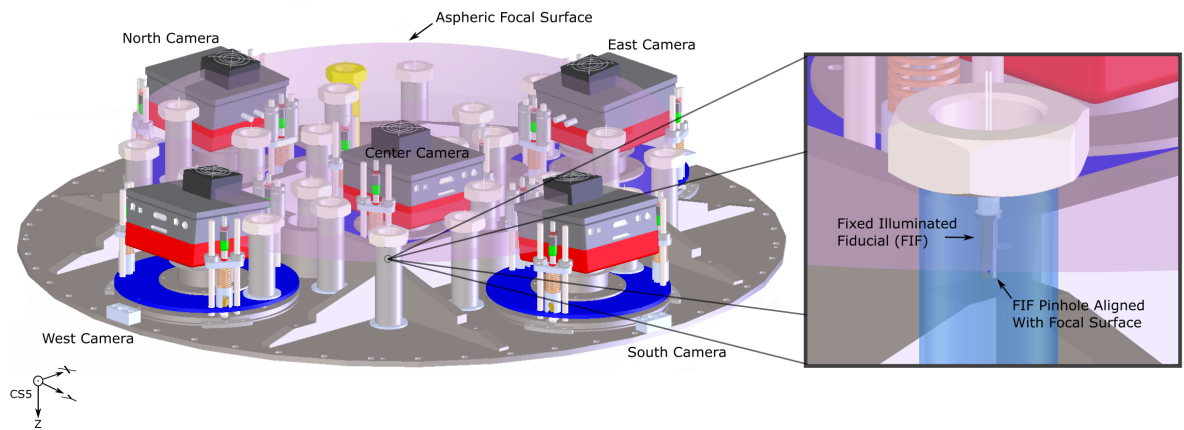

Figure 2. Drawing of the DESI commissioning instrument with aspheric focal surface overlay. The enlarged image on the right shows a fixed illuminated fiducial (IF) inside of its mount and properly aligned such that its illuminated pinhole lies on the focal surface. Proper alignment of the IFs will ensure that they are in focus when imaged by the fiber viewing camera (FVC)

Before we measure the full CI, we will individually align the CCDs on their TTFs. This will reduce the number of major adjustments that will be necessary during the full CI metrology procedure. However, the positions of all of the cameras and IFs will be verified during the final CI measurement run.

### 3.1 Methodology

To perform the metrology measurements, we will attach a custom designed DESI metrology microscope (DMM) to a coordinate measuring machine (CMM) and image the camera and IFs on the CI. The DMM, shown in Fig. 3, combines two parfocal measurement devices: a $100\mu m$ LED illuminated pinhole projector and an imaging CCD. We use the $100\mu m$ pinhole to project light onto the surface of the camera CCDs and measure their tip, tilt, and focus, or image a IF.

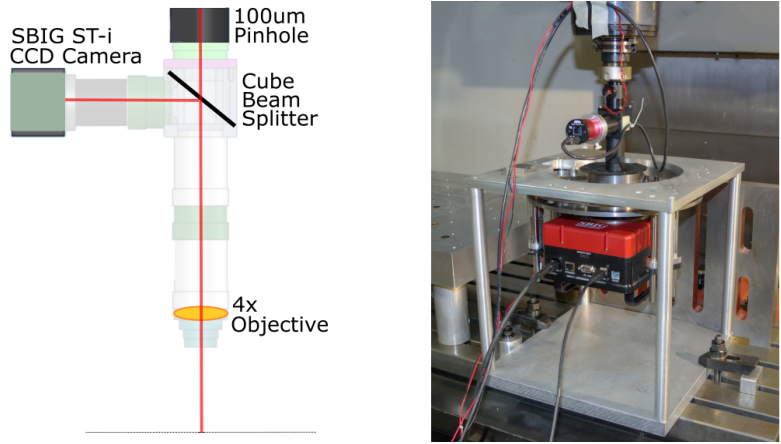

Figure 3. Left: A diagram of the DESI metrology microscope (DMM), comprised of two parfocal measurement devices: a 100$\mu$m LED illuminated pinhole projector and a SBIG ST-i imaging CCD camera, separated by a beam splitter. Right: The DMM attached to a Haas CNC while being used to measure the tip, tilt, and focus of a camera on a TTF, as described in Sec. 3.3.1.

### 3.1.1 Parfocalization

We focused the DMM such that the CCD camera is parfocal with the 100$\mu$m pinhole, which we used to project onto the CI CCDs, then focused the CCD camera and the 100$\mu$m pinhole indirectly since they can't directly see each other.

We began by focusing the 100$\mu$m pinhole by illuminating it through the DMM objective with a bright LED shining onto white paper through the DMM 4X objective, then placed a IF in front of the DMM and aligned the IF pinhole to the 100$\mu$m pinhole so that light passes through the pinhole to the lab CCD camera, as shown in Fig. 4, and focused the IF through the 100$\mu$m pinhole onto the lab CCD. This parfocalizes the IF with the 100$\mu$m pinhole. We then adjusted the height of the DMM camera on the DMM body such that the IF image is in focus. This parfocalizes the 100$\mu$m pinhole with the DMM camera.

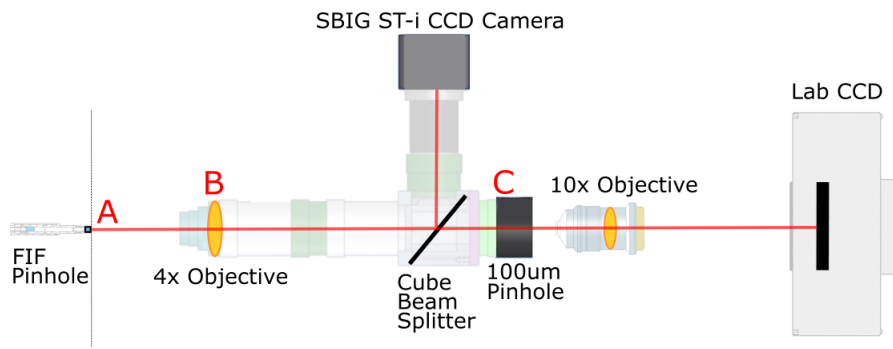

Figure 4. To parfocalize the SBIG ST-i camera relative to the 100$\mu$m pinhole, we performed a DMM calibration procedure that involves co-aligning an IF pinhole (A), the 4X microscope objective axis (B), and the 100$\mu$m pinhole (C). We centered the image of the IF pinhole on the 100$\mu$m pinhole by magnifing the image of the IF pinhole superimposed on the 100$\mu$m pinhole projected onto a lab CCD with a 10X microscope objective, and analyzing the images for concentricity. We then adjusted the height of the DMM camera on the DMM body such that the IF image that it images is in focus, thus parfocalizing the 100$\mu$m pinhole with the DMM SBIG ST-i camera.

## 3.2 DESI Commissioning Instrument Metrology Software

The DESI CI metrology software analyzes input images of IFs or camera CCDs and guides us in adjusting the IFs or cameras such that they are properly aligned with the DESI aspheric focal plane. To make height adjustments in the CS5 Z direction, the software accepts images of a IF or the DMM 100$\mu$m pinhole that was projected onto the TTF camera's surface, and creates a focus curve that solves for height adjustments. An example is shown in Fig. 7. To guide our adjustment of the tip or tilt of the TTF, the DESI CI metrology software accepts images of the DMM 100$\mu$m pinhole projected onto the TTF camera's surface at three locations that form the points of an equilateral triangle, and generates focus curves for each point that determine the tip and tilt of the camera relative to nominal values for those points.

To adjust the TTFs and IFs on the CI in the CS5 X-Y plane, the DESI CI metrology software uses centroiding routines to locate a IF, or light from the 100$\mu$m pinhole projected onto a camera CCD, in an image and find its position relative to the CS5 origin compared to the nominal position.

## 3.3 Initial Alignment

We initally align the TTFs with the DMM attached to a computer numeric controlled (CNC) machine located in The Ohio State University Department of Astronomy, and image the camera and associated IF for each TTF at various locations. We then use the DESI CI metrology software to determine the positions of the TTF instruments in the coordinate system for the FPS. Fig. 5 is a drawing of one of the five TTFs.

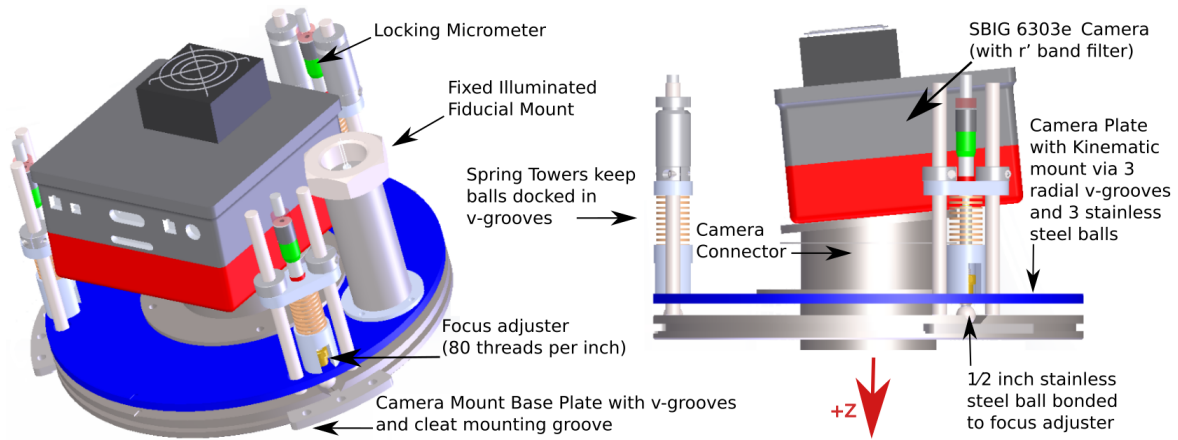

Figure 5. There will be five Tip/Tilt/Focus (TTF) mounts on the CI. Each TTF will have installed a camera and IF, and can be adjusted such that the camera CCDs and IF illuminated pinholes are aligned on DESI aspheric focal surface. The camera is attached to the TTF plate by the camera connector tube. The camera connector tubes are tilted for the north, east, south, and west cameras, whereas the center tube is straight; this is so that the cameras can be more easily adjusted to the contour of the focal surface, as shown in Fig. 2. The TTF kinematic mount has three locking micrometers that are connected to steel balls. Each ball sits in a radial v-groove that is attached to the TTF base plate. By adjusting the micrometers, we tilt or tip the camera plate, thereby adjusting the position of the camera. The IF mount, however, is attached to the TTF base plate, not the camera plate, so to adjust the height of the IF we adjust the threaded base of the IF in the IF mount.

### 3.3.1 Camera Alignment on Tip/Tilt/Focus Mount

To adjust the tip, tilt, and focus of a camera on a TTF, we project light through the DMM's 100$\mu$m pinhole onto the surface of the camera's CCD at three points that form the vertices of an equilateral triangle. We image the projected pinhole and use the DESI CI metrology software, to create a focus curve for each point on the triangle. We then use the focus curves to adjust the tip, tilt, and focus of the camera.

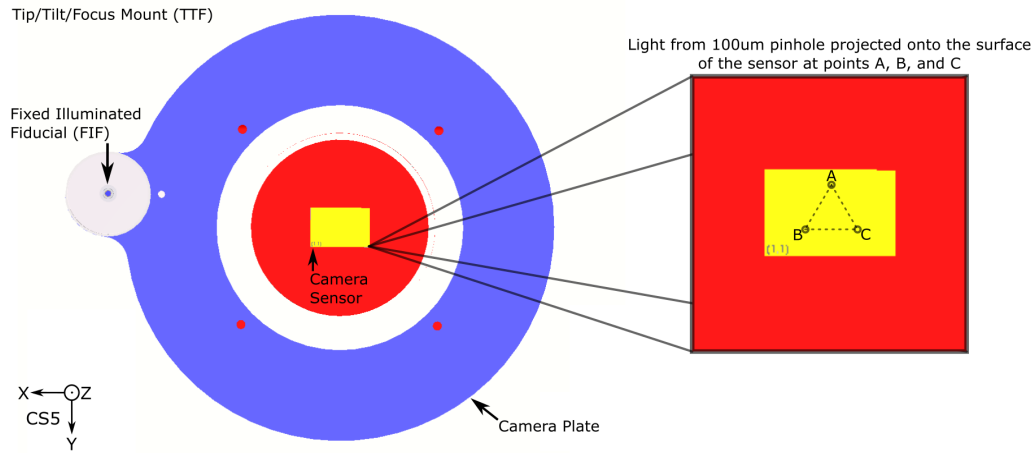

Figure 6. To adjust the tip, tilt, and focus position of the camera on a TTF, we image light from a pinhole that we project onto the camera CCD's surface (shown in yellow). We project the light from the $100\mu m$ pinhole at three locations that form the points of an equilateral triangle, labeled "A", "B", and "C" in the figure. The focus curves that we generate for each point allow us to determine any adjustments to the position of the camera that are needed to align it with the DESI aspheric focal plane.

### 3.3.2 Illuminated Fiducial Alignment

The IF for each TTF sits inside of a IF mount (see Fig. 5). To align the IF to the aspheric focal surface, we image the IF at various points and use the DESI CI Metrology software to construct a focus curve that guides our height adjustments. Fig. 7 shows an example of a focus curve generated by the DESI CI Metrology software. We create the focus curves for both the camera alignment and IF alignment using the standard deviation method, where we find the standard deviation of the light levels in each of the images, to find the desired height for the IF.[3] We will also align the 17 other IFs using the same procedure.

### 3.4 Metrology

There is one IF of each of the five TTFs, as well as 17 IFs spread across the CI focal surface. The relative positions of all 22 Ifs must be measured within $10\mu m$ accuracy within a plane parallel to the focal surface. Since the DESI focal plane assembly will contain IFs that will be imaged by the FVC and used to identify the positions of any misplaced fibers that need to be corrected by the fiber positioners, we use the IFs on the CI to ensure that IFs which are aligned on the DESI aspheric focal plane will be imaged by the FVC. The IFs are arranged on the CI to span a range of radial distances from the CS5 (X,Y) origin. Fig. 8 shows the arrangement of IF on the CI. For the IFs that are mounted on the TTFs, the relative position of the camera CCD and IF are measured with $5\mu m$ accuracy within a plane parallel to the focal surface.

The metrology procedure for the full CI will be performed using the DMM attached to a coordinate measuring machine (CMM), that is large enough to measure the entire volume of the DESI CI focal plate. To mount the CI on the CMM, we will attach the CI to an adapter cylinder that is parallel to the CMM base, thus establishing the CS5 (X, Y) plane. We will then verify the parallelism using a conventional CMM probe. To establish the CS5 Z axis we will locate the center of the adapter cylinder by using the DMM to locate and focus on two illuminated dowel pins that each sit at known locations.

Tab. 1 lists our expected error sources and summarizes the error budget. To further reduce measurement errors in the DESI CI metrology procedure, we will perform all of our alignment and metrology measurements multiple times to reduce the uncertainty.

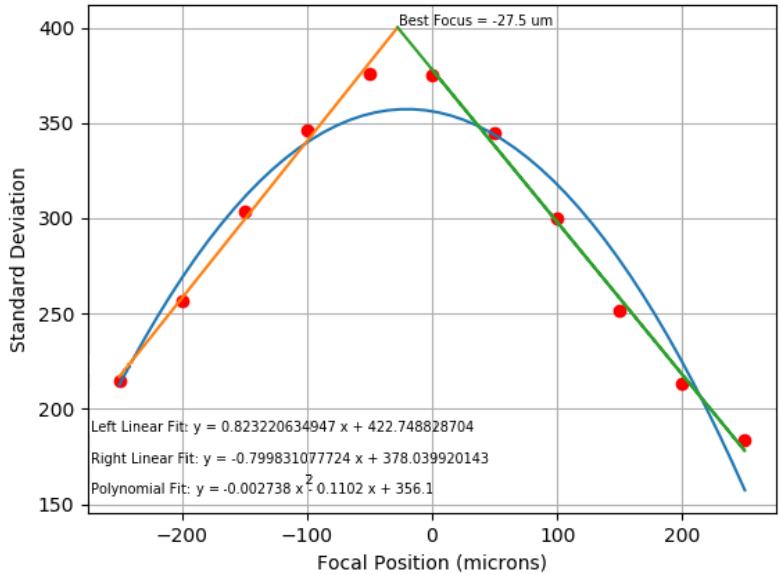

Figure 7. A focus curve for a DESI IF mounted on a TTF. To collect the information needed to create a IF focus curve we image the IF with the DMM positioned at various CS5 Z heights. The DESI CI metrology software then finds the standard deviation of the light level in each image, and applies a polynomial fit to the data (blue curve). We use the focal position where the derivative of the polynomial fit is zero to split the data into "left" and "right" groups, and apply a linear fit to each side. the point where the linear fits intersect is the the focus position for the given IF relative to our visual estimate of the best focus position. In this example, the calculated best focus in CS5 Z is -27.5$\mu$m away from our visual estimate, indicating that we need to lower the IF to ensure that it is aligned with the DESI aspheric focal plane.

Table 1. Summary of metrology error budgets for CI CCD cameras and IFs.

| CCD metrology | CMM single axis single measurement error | +/- 4$\mu$m. |
|---|---|---|
| | 100$\mu$m pinhole centroiding error | +/- 1$\mu$m. |
| | CS5 Z (focus) | +/- 5$\mu$m. |
| | RMS CS5 X and CS5 Y error | +/- 2-3$\mu$m. |
| IF metrology | CMM single axis single measurement error | +/- 4$\mu$m |
| | IF pinhole centroiding error | +/- 1$\mu$m. |
| | CS5 Z (focus) | +/- 5$\mu$m. |
| | RMS CS5 X and CS5 Y error | +/- 2-3$\mu$m. |
| CCD to nearest IF relative metrology | RMS CS5 X and CS5 Y error | +/- 2-3$\mu$m. |
| | DMM calibration | +/- 1-2$\mu$m. |

## 4. SUMMARY AND PLANS

While we have described the procedure for aligning and performing the metrology of the DESI CI, we have also shown that our preliminary measurements have been successful. All of the CI components will be aligned with the DESI focal surface and measured with high precision. The resulting data will be used by the PlateMaker software to map locations on the focal plane with celestial objects using an accurate model of the distortion pattern created by the primary mirror and optical corrector, and the distortion pattern over the DESI field of view as seen by the FVC caused by the optical corrector and FVC optics. To date we have successfully measured the individual TTFs, and will soon have all of the TTFs pre-aligned and ready to be installed on the CI. In the following months, we plan to finish our TTF pre-alignment using the Haas CNC at The Ohio State University, and install them, and the additional 17 IFs, on the CI. We will then measure the entire CI focal plate with a CMM and ship the CI to the Mayall telescope for commissioning late this year.

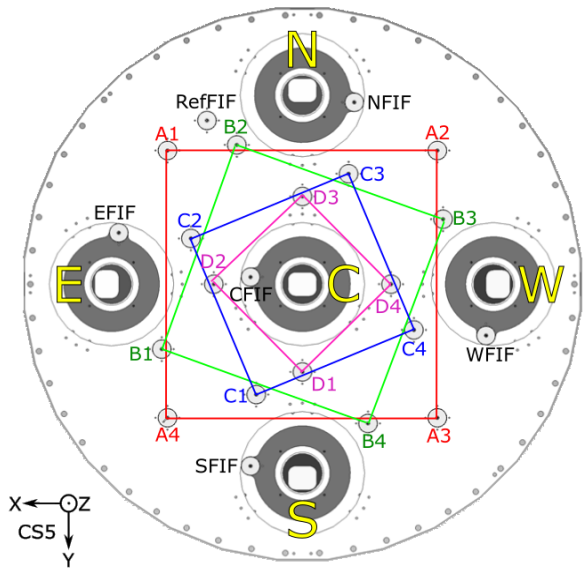

Figure 8. The locations of the IF on the DESI CI are grouped such that: each TTF has one IF, four groups of four IFs sit at four different radii as measured from the CS5 (X,Y) origin, and an additional reference IF sits at its own unique radii. The reference IF, labeled "RefFIF" in the diagram, we use identify the orientation of the CI on DESI since all of the other instruments on the CI are radially symmetric. The distance between IF within a IF group is equal, forming a square that is labeled here as A, B, C, and D.

## ACKNOWLEDGMENTS

We would like to thank Jonathan Shover from The Ohio State University Department of Astronomy machine shop for his tireless efforts in machining commissioning instrument components. In addition, we would like to thank Daniel Pappalardo from The Ohio State University Department of Astronomy for his assistance in setting up the CI and DMM electronics.

This research is supported by the Director, Office of Science, Office of High Energy Physics of the U.S. Department of Energy under Contract No. DEAC0205CH1123, and by the National Energy Research Scientific Computing Center, a DOE Office of Science User Facility under the same contract; additional support for DESI is provided by the U.S. National Science Foundation, Division of Astronomical Sciences under Contract No. AST-0950945 to the National Optical Astronomy Observatory; the Science and Technologies Facilities Council of the United Kingdom; the Gordon and Betty Moore Foundation; the Heising-Simons Foundation; the National Council of Science and Technology of Mexico, and by the DESI Member Institutions. The authors are honored to be permitted to conduct astronomical research on Iolkam Du'ag (Kitt Peak), a mountain with particular significance to the Tohono O'odham Nation.